# Asymmetric Electron Transmission across Asymmetric Alkanethiol Bilayer Junctions

Michael Galperin and Abraham Nitzan

*School of Chemistry, the Sackler Faculty of Sciences, Tel Aviv University, Tel Aviv, 69978, Israel*

Slawomir Sek[*] and Marcin Majda

*Department of Chemistry, University of California, Berkeley*

*Berkeley, California 94720-1460, U.S.A.*

## Abstract

Asymmetric i/V curves with respect to the polarity of the voltage bias were observed in the Hg-Au junctions containing bilayers of alkanethiols of different chain length. Larger current resulted when a negative bias was applied to the metal carrying a longer chain alkanethiol monolayer. This behavior is simulated using a single molecule junction model, within the frameworks of the extended Hückel (EH) model and the nonequilibrium Green's function formalism at the Hartree Fock level (NEGF-HF). Qualitative agreement with the experimental results with respect to the magnitude and sign of this asymmetry is obtained. On the basis of the NEGF-HF calculation, the origin of the effect is suggested to be the asymmetric behavior of the character of the junction highest occupied molecular orbital (HOMO) at opposite biases. This change of character leads to different effective barriers for electron transfer for opposite signs of the voltage drop across the junction.

[*] Permanent address: Department of Chemistry, Warsaw University, Pasteura 1, 02-093 Warsaw, Poland

## 1. Introduction

An important attribute of molecular junctions is the behavior of the current under reversal of the voltage bias. Current rectification, loosely defined as a phenomenon in which the magnitude of the junction current depends on the polarity of the voltage bias, has been observed in a number of experiments involving monolayer and multilayer films as well as single molecule type junctions and STM measurements.[1] We present a set of experimental results obtained with the Hg-Au junctions incorporating a bilayer of alkanethiol films of variable chain length. These measurements revealed that current rectification was observed for all asymmetric junctions, or those in which a monolayer assembled on one metal surface consisted of longer chain molecules than those formed on the other metal surface before a junction was assembled. In such cases, a larger current was measured when the polarity of a voltage bias directed electron tunneling first through a longer alkanethiolate segment of the bilayer. This is illustrated schematically in Figure 1 where a larger current would be observed if the left hand side of the junctions were biased negative with respect to its right hand side. Below, we present the description of the Hg-Au junctions and address the asymmetry of the i/V curves using a simple molecular model.

The methodology used in the formation of the Hg-Au junctions is analogous to the one we developed for the Hg-Hg type junctions.[2, 3] The use of mercury in these junctions is also similar to the Hg-Ag junctions described by Whitesides,[4, 5] to Hg-SiO$_2$-p-Si junctions of Cahen,[6] and to Hg-C junctions reported recently by McCreery.[7] An overview of the related research with scanning probe microscopy can be found in a recent report by Bard and co-workers.[8] In this report, in order to improve the stability of the junctions and reproducibility of the resulting *I/V* curves, we first characterize the structure of the alkanethiol monolayers on gold substrates by recording standard cyclic voltammetric current voltage curves in aqueous electrolytes of a redox probe. On the basis of these measurements, only those monolayer-coated gold substrates exhibiting no pin-hole defects are selected and used in the subsequent junction experiments.

The first prediction of molecular rectification is due to Aviram and Ratner,[9] where a simple model for molecular electronic structure which provides rectification was proposed based on analogy to rectifying p-n junctions in solid-state devices. Later studies[10-13] have discussed different scenarios for molecular rectifiers. In particular, Waldeck and Beratan[10] discuss the possibility of the asymmetry in the *width* of the tunneling barrier as a function of applied voltage, resulting from the interplay between the positions of the molecular HOMO or LUMO relative to the leads Fermi energies and their voltage induced shifts. Other works[11-13] emphasize the importance of the potential profile on the molecule for the rectifying behavior of the junction. Datta and coworkers[11] have emphasized the effect of the voltage division factor, a measure of how the voltage drop across the junction is divided between the two metal-molecule contacts, on the rectification behavior, while Mujica, Ratner, and Nitzan[13] have argued that in many situations where the voltage drops take place entirely at these contacts, irrespective of the voltage division factor $\eta$, no significant rectification will occur. The latter authors argue that in addition to the essential structural asymmetry, the existence of voltage drop *along* the molecular bridge is an important attribute of a rectifying molecular junction. In this respect we note that a significant voltage drop along an essentially 1-dimensional bridge is expected from electrostatic considerations even for highly polarizable molecular chains.[14, 15] This potential distribution can be in principle controlled, for example by introducing weak links into the structure as discussed below and in Refs. [4, 5]

This work presents experimental results that show *i*/*V* asymmetry in junctions similar in character to those described in [4, 5], however with very simple molecular bridges, made of alkanethiol bilayers with structural asymmetry derived from molecular components of different alkyl chain lengths. We also present a theoretical analysis of these observations. Our theoretical treatment extends previous studies in three aspects. First, we take into account the change in the electronic structure of the bridge when an external potential bias is applied to the system. Second, we take into account electron-electron interactions on the molecule, describing the electrodes as reservoirs of free carriers at equilibrium. Finally, while some insight may be obtained from simple independent electron models such as the extended Hückel approximation, we base our main considerations on the non-equilibrium Green's function method, which provides a powerful tool for investigations of many-body quantum systems open to particle and energy fluxes.

The next two sections describe our experimental methodology and the experimental results. In Section 4 we introduce our theoretical model and method. Section 5 presents and discusses our theoretical results. Section 6 concludes.

## 2. Experimental

**Reagents**. Fresh samples of alkanethiols, $C_nSH$ (n=9, 10, 12 and 16) were purchased from TCI America (95+%) or from Aldrich. Hexadecane (Aldrich, 99+%), mercury (Quicksilver, Inc., triply distilled), $Ru(NH_3)_6Cl_3$ (99%, Strem), KCl (Fisher, ACS grade), and the alkanethiols were used as received. House distilled water was passed through a 4-cartridge Barnstead Nanopure II purification system. The resistivity of the final product was 17.8 – 18.3 MΩ cm.

**Gold-coated glass substrates with self-assembled alkanethiols monolayers**. The gold-coated glass substrates were produced by vacuum (1-5 x $10^{-7}$ Torr) vapor-deposition of Cr (2-5 nm) and Au (80-100 nm) films on glass slides which were then cut into 0.8 x 2.0 $cm^2$ sections. Self-assembly of alkanethiol monolayers involved incubation of the freshly produced gold-coated substrates in 1 mM ethanol solutions of an alkanethiol for ca 24 hr. The monolayer-coated gold substrates were used in the electrochemical experiments within 2-3 days.

**Hg-Au tunneling junctions**. The mercury-gold junctions incorporating alkanethiolate bilayers were assembled in a junction apparatus described in a previous report.[3] It consisted of a micrometrically driven Kemula-Kublik type hanging mercury drop electrode (HMDE), an x-y-z micrometer stage (Model M-460A Newport Corp.) supporting a gold-coated glass slide, and a long focal length microscope (Titan, Tool Supply Co. Inc.) affording a 100 fold magnification and equipped with a filar eye piece. The HMDE was mounted vertically with a tip of a 250 μm capillary pointing upwards. A gold-coated glass substrate was attached (with a double-stick tape) to a vertically oriented aluminum rod with the gold-coated surface facing downwards. The position of the rod and thus of the gold-coated glass slide was controlled with the x-y-z micrometer. Prior to a junction assembly, a small volume of a hexadecane solution of a selected alkanethiol (typically 10 % v/v) was placed on the circular cross section of a HMDE glass capillary. Next, when a Hg drop (500 – 800 μm in diameter) was generated, its entire volume remained in the hexadecane solution while its apex was approximately coplanar with the surface of the small hexadecane pool. Our earlier experiments with alkanethiols assembly on the expanding Hg drop electrodes assure us that the procedure just outlined results in a formation of a mercury alkanethiolate monolayer on the Hg surface.[16] A dry, gold-coated glass substrate coated with a self-assembled monolayer of a selected alkanethiol (often different than that used to form a monolayer on mercury) was subsequently lowered to contact the Hg drop. An effort was made not to generate any pressure between the gold and mercury surfaces. A twitch phenomenon which was described earlier was usually observed within seconds after a visible contact between the Hg and Au surfaces was first established.[3] As previously, we associate this with removal of the remaining solvent from the junction area and thus establishment of a bilayer junction.[3] This phenomenon is also associated with a slight enlargement of the Au-Hg contact area. The length of the contact line between the two metals is then measured microscopically (with ca 5-10% precision) typically yielding the junction area of $2 \times 10^{-3}$ cm$^2$.

**Electrochemical and tunneling current measurements**. Recording of the current – voltage bias curves (i/V) and electrochemical characterization were carried out with a CHI Model 660A electrochemical analyzer (CH Instruments, Inc. Austin, TX) in three- or two-electrode, potentiostatic configuration, respectively.

## 3. Experimental results

Assembly of the Hg-Au bilayer junctions involved two separate experimental stages. The first consisted of self-assembly of alkanethiol monolayers on gold substrates and their electrochemical characterization. The second and critical stage involved formation of the alkanethiolate monolayer on a Hg drop immediately followed by assembly of Hg-S-C$_n$/C$_m$-S-Au junctions and recording of the i/V curves. We introduced the first stage in order to improve the stability of the junctions and reproducibility of the i/V measurements.

Following self assembly of the alkanethiol monolayers ($C_9$, $C_{10}$, $C_{12}$, see Experimental Section), individual monolayer-coated electrodes were immersed in 1.0 mM $Ru(NH_3)_6^{3+}$, 0.5 M KCl solution to carry out their cyclic voltammetric characterization. In spite of identical treatment of all gold substrates, three types of i vs E curves were obtained in these experiments as shown in Figure 2. The shape of curve A clearly indicates that these electrodes are coated with monolayers populated with a large density of pin-holes defects. The sigmoidal shape of curve B is indicative of a smaller density of such pinholes, each generating a hemi-spherical diffusion zone that does not overlap with those of the neighboring pin-hole defects. The exponentially rising current of curve C can in turn be interpreted as due solely to electron tunneling across an essentially pin-hole free alkanethiol monolayer. The plot of the logarithm of the current measured at -0.45 V vs number of methylene groups in the chain for the type-C electrodes coated with nonane-, decane- or dodecanethiol monolayers was linear with a slope of 1.04 ± 0.1 per $CH_2$ group, consistent with our earlier measurements[16] and those in other reports.[17]

The stability of the Hg-Au bilayer junctions correlated well with the voltammetric behavior of Figure 2. The type A gold substrates did not allow formation of stable junction at all, or generated a short (amalgamation of gold) immediately upon application of a any voltage bias. Junctions assembled between B-type monolayer-coated gold substrates and the alkanethiolate-coated mercury drops were more stable allowing application of 0.7 – 0.9 V bias before breakdown. When the pin-hole free (type C) gold substrates were used to form the Hg-Au bilayer junctions, the breakdown bias extended to ca 2.0 – 2.2 V. In view of these results, only gold electrodes exhibiting C-type behavior were used in the Hg-Au junction experiments. 80% of all monolayer-coated gold substrates did not meet this rather stringent selection criterion and were rejected. However, each of the gold substrates that passed this screening test performed well in the junction experiments producing quantitative data. Since the latter are by far more difficult and time consuming, we consider this screening test a valuable element of the whole protocol.

The exact methodology of junction formation is given in the Experimental Section. All i/V curves (typically ± 1.0 – 1.5 V) were recorded at 500 V/s to minimize the time of exposure of the junctions to high biases. In nearly all cases, the i/V curves were initially superimposable. With time (after 1-2 min), currents tend to increase and ultimately junctions collapse. Since the slow current increase with time is interpreted as a result of slowly progressing deformation of the bilayer structure involving generation of gauche-defects and intercalation of chains,[3] only the i/V curves recorded early on in the life time of each junction were used in data analysis. An example of the i/V curves obtained with different Hg-Au bilayer junctions is shown in Figure 3. A higher current at a negative bias corresponding to the negative voltage being applied to the Hg relative to the Au side of the junction is characteristic for all the asymmetric junctions. Data analysis of the i/V curves such as those in Figure 3 involved determination of the current magnitude at a fixed bias. Logarithm of the average current measured at +1.0 and -1.0 V was plotted vs number of methylene units for all the junctions (see the list in Table 1) and yielded a linear plot with a slope of 0.95 ± 0.06 per $CH_2$. This value is smaller than that we obtained in the experiments involving Hg-Hg junctions of 1.29 ± 0.03 (over a narrower range of 18-24 $CH_2$ groups), [3] but similar to the value reported by Whitesides for the Hg-Ag bilayer junctions of 1.07 ± 0.1 (recorded over a range of 24-32 methylene groups).[18] The ratios of the currents at -1.0 to those at +1.0 V for these junctions are listed in Table 1. This ratio appears to increase with the level of the structural asymmetry of the bilayer junctions. The fact that the current ratio is 1.0 for the symmetric junctions suggests that the small current rectification documented in Table 1 is not introduced merely by the use of two different metals and thus by their different physical and chemical properties such as their work function and the strength of the metal-sulfur bond. [19]

Table 1. Experimental results of the i/V asymmetry in the Hg-Au bilayer junctions.

| Junction | Ratio of current: i(-1.0 V)/i(+1.0 V)[a,b] |
|---|---|
| Hg-SC$_{12}$/C$_{12}$S-Au | 0.98 ± 0.13 |
| Hg-SC$_{12}$/C$_{10}$S-Au | 1.03 ± 0.07 |
| Hg-SC$_{16}$/C$_{12}$S-Au | 1.22 ± 0.16 |
| Hg-SC$_{12}$/C$_9$S-Au | 1.44 ± 0.20 |
| Hg-SC$_{16}$/C$_{10}$S-Au | 1.34 ± 0.19 |
| Hg-SC$_{16}$/C$_9$S-Au | 2.03 ± 0.27 |

[a]Current at the negative bias refers to the measurement with the Hg side of the junction biased negative relative to the Au side.
[b]The current ratios represent averages and standard deviations of 6 to 8 measurements each of which involved an individually assembled junction with a new monolayer-coated gold substrate.

## 4. Theoretical approach

As described above, the junctions are made of two planar metal electrodes covered by unimolecular alkanethiol films that are brought into physical contact without formation of a chemical bond. As a simple model of this system we consider a junction consisting of just two alkanethiol chains, each connected to its corresponding electrode through the thiol group. The chains are placed perpendicular to the (planar parallel) electrode

10surfaces,[*] and face each other with their carbon ends separated by an arbitrarily chosen distance of 2Å (see Fig. 1b). We represent this junction schematically by L-S - $(CH_2)_nCH_3 \ldots CH_3(CH_2)_m$ - S-R, where L and R denote the left and right electrodes and ... is the van-der-Waals gap between the molecular species. In the calculations presented below we consider two types of such junctions: symmetric ($n = m = 3$) and asymmetric ($n = 5, m = 1$). Thus, L corresponds to the Hg and R to the Au electrodes in the experimental investigations. The electrodes are treated as semi-infinite reservoirs of free electrons at thermal equilibrium, characterized by their electrochemical potentials $\mu_L = E_F - eV_L$ and $\mu_R = E_F - eV_R$ (e denotes the absolute value of the electron charge). We will refer to the situation where $\mu_R - \mu_L = e(V_L - V_R) \equiv e\Delta V > 0$, i.e. when the left electrode is biased positively, as the positively biased junction. This model is used below to elucidate the essential features needed to observe asymmetry in the current–voltage characteristic as seen experimentally.

Our approach is based on the Landauer formula[20-23] in the framework of the Keldysh non-equilibrium Green's function (NEGF) method[24] [22] computed on the level of the restricted Hartree Fock (HF) approximation. The core Hamiltonian and the two center integrals are obtained from GAUSSIAN[25], using the LANL2DZ basis set for the atomic orbitals. In order to conform with the bounds imposed by the use of RHF we have performed our calculations on the closed shell molecular structures L-HS - $(CH_2)_nCH_3 \ldots CH_3(CH_2)_m$ - SH-R, and have assigned the self energies associated with the molecule–metal interactions (Eq. (3) below) to the terminal S atoms. The Löwdin procedure[26] is used to transform to an orthonormal basis. In this basis the

---

[*] This is a simplification, and in general other orientations need to be considered, e.g., alkanethiols adsorb on gold with a tilt angle of $30^0$.

Hamiltonian of our open quantum system becomes in the second quantization representation

$$\begin{aligned} H &= \sum_{i,j;\sigma} H^c_{ij} c^\dagger_{i;\sigma} c_{j;\sigma} + \sum_{i,j;\sigma} V^{ext}_{ij} c^\dagger_{i;\sigma} c_{j;\sigma} \\ &+ \sum_{i_1,i_2,i_3,i_4;\sigma} V^{i_1 i_2}_{i_3 i_4} c^\dagger_{i_1;\sigma} c^\dagger_{i_2;\sigma} c_{i_4;\sigma} c_{i_3;\sigma} \\ &+ \sum_{k\in L,R;\sigma} \varepsilon_k c^\dagger_{k;\sigma} c_{k;\sigma} \\ &+ \sum_{k\in L,R;\sigma} \left( V_{ki} c^\dagger_{k;\sigma} c_{i;\sigma} + h.c. \right) \end{aligned} \quad (1)$$

where the $i$ and $j$ labels are used for the Löwdin (orthogonalized atomic) orbitals of the molecule, $k$ denotes (free) electron states in the contacts L and R and $\sigma =\uparrow,\downarrow$ is the spin label. Here and below we use Greek indices to indicate atomic orbitals on the molecule while Latin indices denote Löwdin orbitals. $c$ and $c^\dagger$ are annihilation and creation operators of an electron in the corresponding orbital. The first two terms in Eq. (1) are the single electron part of the molecular Hamiltonian and the external voltage drop along the junction. For planar parallel electrodes the external potential is a linear function of the distance between the electrodes, i.e., if D is the distance between the metal planes and $x=0$ is the position of the left plane, then $V_{ext}(x) = V_L - (x/D)\Delta V$ where $\Delta V = V_L - V_R$. The matrix $V^{ext}_{ij}$ in the Hamiltonian (1) is the Löwdin transform of the matrix $V^{ext}_{\alpha,\alpha'} = \langle \alpha | V_{ext} | \alpha' \rangle$ in the atomic orbital representation. Note that the metal electrodes are assumed to be in equilibrium characterized by the electrochemical potentials µ$_L$ and µ$_R$ respectively. The third term in Eq. (1) represents the electron-electron interaction on the molecule. The matrix elements $H^c_{ij}$ and the tensor elements $V^{i_1 i_2}_{i_3 i_4}$ (essentially two-center integrals) are obtained from the quantum calculation done on the isolated molecule using the Gaussian package. The last two terms in Eq. (1) represent free electrons in the electrodes and the molecule-electrode coupling.

The Landuaer formula, applied to the evaluation of the current dependence on the applied voltage leads to[22, 27]

$$I = -\frac{e}{\hbar} \int_{-\infty}^{\infty} \frac{dE}{2\pi} Tr\left[\Gamma_L(E) G^r(E) \Gamma_R(E) G^a(E)\right] \left(f_L(E) - f_R(E)\right) \qquad (2)$$

where the trace is taken over the subspace of the molecular bridge,

$$\Gamma_K(E) = i\left[\Sigma_K^r(E) - \Sigma_K^a(E)\right] \quad ; \quad K = L, R \qquad (3)$$

is (without the factor $i$) the anti-hermitian part of the self-energy (see below) incurred by reducing the description of the process to the molecular subspace and where $f_K$ ($K=L,R$) are the Fermi functions of the electrodes,

$$f_K(E) = \left[e^{(E-\mu_K)/k_B T} + 1\right]^{-1} \qquad (4)$$

The self-energy $\Sigma$ is the molecular-subspace expression of the molecule-electrode coupling. In the non-equilibrium (Keldysh) formalism for our model it is given by

$$\left[\Sigma_K\right]_{ij;\sigma}(\tau_1, \tau_2) = \sum_{k \in K} V_{ki}^* V_{kj} \left[G_K\right]_{k;\sigma}(\tau_1, \tau_2) \quad ; \quad K = L, R \qquad (5)$$

where again $i, j$ denote Löwdin orbitals on the molecular bridge, $k$ labels the free electron states on the electrodes and σ specifies the spin state. $G_K(\tau_1, \tau_2)$ is the contour ordered (Keldysh contour) Green function of the electrode $K$, with $\tau$ denoting the contour time variable.[24, 28] Its projections on the real time axis are the retarded and advanced, $\Sigma^r = (\Sigma^a)^\dagger$, lesser, $\Sigma^<$ and greater $\Sigma^>$ self-energies (see, e.g., [29, 30]). For semi-infinite free electron metals or for metals represented by a tight-binding model with non-interacting electrons $\Sigma^r$ may be calculated exactly. However, since we focus here on qualitative effects of the molecular electronic structure rather than on a quantitative assessment of the influence of the molecule-metal contacts on the current–voltage characteristic, we limit ourselves to a simple model for the self-energy, disregarding its real part and taking its imaginary part in the atomic orbital representation to be diagonal, and non-zero only for orbitals belonging to the sulphur atoms. For these orbitals we assume the wide band limit for the self-energy, i.e., taking a constant (energy independent) width arbitrarily chosen in our model as $\gamma=0.2$eV (an order of magnitude estimate of the inverse lifetime for the decay of an excess electron on the S atom into the continuum of metal electronic states if unoccupied). With this simplification, the projections of the self-energy onto the real time axis in the local atomic basis representation become (after Fourier transforming to energy space)

$$
\begin{aligned}
\left[\Sigma_K^r\right]_{\alpha\beta;\sigma} &= -i\delta_{\alpha,\beta}\frac{\gamma}{2}\bigg|_{\alpha\in S_K} \\
\Sigma_K^a &= \left[\Sigma_K^r\right]^\dagger \\
\left[\Sigma_K^<\right]_{\alpha\beta;\sigma} &= i\delta_{\alpha,\beta}f_K(E)\gamma\bigg|_{\alpha\in S_K} \\
\left[\Sigma_K^>\right]_{\alpha\beta;\sigma} &= -i\delta_{\alpha,\beta}(1-f_K(E))\gamma\bigg|_{\alpha\in S_K}
\end{aligned} \quad ; \quad K=L,R \qquad (6)
$$

where $S_K$ denotes the left ($K=L$) or right ($K=R$) sulphur atoms and where $\alpha$ and $\beta$ are atomic orbitals. The corresponding $[\Sigma_{L/R}]_{ij;\sigma}(E)$ are the Löwdin transformations of these matrices.

The electron-electron interaction, third term in Eq. (1), can be taken into account by expanding the contour ordered $S$-matrix[22, 24] in increasing orders of this interaction. We restrict ourselves to the lowest non-vanishing term (second order), which is essentially the Hartree-Fock (HF) approximation. The self-energy for this case reads

$$
[\Sigma_{HF}]_{ij;\uparrow} = \sum_{l,m}\left(V_{jm}^{ik}\left[\rho_{ml;\uparrow}+\rho_{ml;\downarrow}\right]-V_{mj}^{ik}\rho_{ml;\uparrow}\right) \qquad (7)
$$

where $V_{i_3 i_4}^{i_1 i_2}$ are Löwdin transforms of

$$V_{\alpha_3\alpha_4}^{\alpha_1\alpha_2} = \int d\vec{r_1} \int d\vec{r_2} \phi_{\alpha_1}^*(\vec{r_1}) \phi_{\alpha_2}^*(\vec{r_2})$$
$$\times \frac{e^2}{|\vec{r_1}-\vec{r_2}|} \phi_{\alpha_3}(\vec{r_1}) \phi_{\alpha_4}(\vec{r_2}) \qquad (8)$$

and the summation is over the molecular subspace. Similar equation holds for the opposite spin. Here the first term is the Coulomb interaction and the second is the exchange integral. $\rho_{mk;\sigma}$ are elements of the density matrix

$$\rho_{mk;\sigma} = -i \int_{-\infty}^{+\infty} \frac{dE}{2\pi} G_{mk;\sigma}^<(E) \qquad (9)$$

with $G_{mk;\sigma}^<(E)$ being the lesser Green function in the molecular subspace. Although the HF approximation is a mean field theory that does not take into account electron correlation, it appears to be enough for explaining the qualitative behavior of the current–voltage characteristics of molecular junctions. We note in passing that electron correlations could be taken into account by considering higher order terms in perturbative expansion or, semi-empirically by using DFT (LDA) method as is done, e.g. in Ref. [31].

The calculation is done in the standard way, solving simultaneously the Dyson equation for the retarded (or advanced) Green's functions and the Keldysh equation for the lesser/greater projections, Eqs. (10) and (11),

$$G^r(E) = \left[ E - H_0 - \Sigma^r \right]^{-1} \qquad (10)$$

$$G^{<,>}(E) = G^r(E) \Sigma^{<,>}(E) G^a(E) \qquad (11)$$

all defined in the molecular subspace of the problem. $H_0$ stands for the two first terms of the Hamiltonian (1), $G^a(E) = \left[ G^r(E) \right]^\dagger$, and the self-energies are

$$\Sigma^r = \Sigma_L^r + \Sigma_R^r + \Sigma_{HF} \qquad (12)$$

$$\Sigma^{<,>}(E) = \Sigma_L^{<,>}(E) + \Sigma_R^{<,>}(E) \qquad (13)$$

Since the problem is spin-symmetric, it is enough to consider only one spin coordinate, so here and below we omit the spin index. Eqs. (6)-(13) are solved iteratively using the following sequence of steps:

1. For the given values of the electrochemical potentials of the electrodes, $\mu_K$ ($K=L,R$), calculate the self-energies due to the leads from Eqs. (6). The HF self-energy, Eq. (7), is set to be zero at the first step of iteration.

2. Form the retarded, lesser, and greater self-energies of the system using Eqs. (12) and (13).[*]

3. Calculate the retarded, lesser, and greater Green functions from Eqs. (10) and (11).[*]

4. Calculate the density matrix from Eq. (9) and use it to update the HF self-energy according to Eq. (7).

5. Check for convergence. Convergence is achieved when

$$\left| \frac{[\Sigma_{HF}]_{ij}^{(l+1)} - [\Sigma_{HF}]_{ij}^{(l)}}{[\Sigma_{HF}]_{ij}^{(l)}} \right| < \delta \quad (all\ i, j) \tag{14}$$

where $l$ is the iteration step and $\delta$ is a predefined tolerance. If convergence has not yet been achieved return to step 2.

After the procedure has converged, the current through the junction is obtained from the Landauer formula (2). Note, that Eq. (2) is applicable only for coherent transport,[†] and that electron-electron interaction does not destroy coherence on the HF level. In the general case (e.g. when higher order terms in perturbation expansion are taken into

---

[*] It is sufficient in fact to focus on the retarded and lesser entities, keeping in mind that the other two are related to these by $G^a = G^{r\dagger}$ and $G^r - G^a = G^> - G^<$.

[†] In the literature on bridge assisted tunneling, "coherent electron tunneling" is the term used to describe electron tunneling undisturbed by phase destroying thermal motions of the nuclei.

account or if electron–phonon interaction is present) a more general expression for the current should be used (see e.g. [30])

Generally, the problem should be presented on an energy grid large enough to span the energy range relevant for the problem and dense enough for accurate evaluation of the integral in Eq. (9). For large systems this leads to huge demand on memory and CPU time needed for the evaluation of the $G$ and $\Sigma$ matrices (in particular the need to invert, at each iteration, big matrices at each point of the energy grid is very time consuming). At the HF level of theory and when Eq. (6) holds for the self-energy elements associated with the molecule–metal coupling, one can simplify the procedure so that there is no need for a brute force numerical evaluation of the integral in Eq. (9). For details see Appendix A.

## 5. Results and discussion

The calculated current-voltage characteristic depends on the alignment of the molecular levels relative to the metal Fermi energies. In the calculations described below we have arbitrarily chosen the Fermi levels of the unbiased electrodes to be in the middle of the HOMO-LUMO gap of free unseparated alkyl dithiol (HS-$(CH_2)_n$-SH) molecules as computed by the extended Huckel method or by HF for the corresponding calculations described below.[*] The HOMO-LUMO gaps of the separated structures HS-$(CH_2)_n CH_3$ . . . $CH_3(CH_2)_m$-SH are only slightly different. For the HF calculation we have found that different choices of Fermi energies did not influence the qualitative picture presented below.

It is of interest to consider first the behavior predicted for this system by a simpler computational scheme, the extended Hückel method, which does not take into account electron-electron interactions. On this level of the calculation, the voltage drop across the junction cannot be computed and has to be entered as part of the model. This is done (see, e.g. Ref. [11]) by modifying the zero order energy of each atomic orbital in the bridge molecule by the imposed local potential at the corresponding atom. We consider three typical cases: A. The potential drops linearly along the molecular bridge, i.e. $V_{ext}(x) = V_L - (x/D)\Delta V$. B. The potential drops only at the molecule–electrode interface (we take equal drops on the left and right contacts), and remains constant along the bridge itself. C. (for the separated molecules) The potential drops at the molecule–electrode interfaces and at the intermolecular gap (see Fig. 4 for schematic representation; in the present calculation we take equal drops in the three locations).

---

[*] Specifically, $E_F$ was taken -0.07au and -0.15au in the HF and EH calculations, respectively.

Figure 5 shows the results obtained from an extended Huckel calculation for cases A, B, and C for the symmetric ($n=m=3$; dashed line) and asymmetric ($n=5$, $m=1$; full line) junctions. We see that asymmetry in the I/V curve, qualitatively similar to the behavior seen in Fig. 3 and Table I appears in the asymmetric junction in cases A and C but not in case B, demonstrating the sensitivity of the resulting behavior to the choice of potential distribution on the bridge. Obviously the potential distributions used above are arbitrary choices. In addition, charge accumulation at the edges of the structural gap (that effectively constitutes a molecular capacitor) may cause the electronic structure of the molecular bridge to depend on the applied voltage through more than a simple energy shift. To account for such effects electron-electron interactions need to be included in the model, and the NEGF-HF method described in Section 4 is the simplest possible approximation. Next we describe the result of calculations based on this approach.

Figure 6a shows the current–voltage characteristics for the symmetric and asymmetric junctions obtained by the NEGF-HF calculation. The symmetric junction naturally shows a symmetric *I/V* behavior, while the non-symmetric junction shows asymmetry in the current–voltage characteristic such that the current increases more rapidly when the left electrode (with the longer molecular chain) is biased negatively. Again, this behavior is in agreement with the experimental results, now obtained without imposing any particular potential distribution model.[*] As is seen in Fig. 6b, the observed asymmetry in the *I/V* behavior increases with increasing asymmetry in the molecular structure, consistently with the trend in the experimental results (see Figure 3).

The behavior observed in the I/V characteristics of these junctions can be explained by looking closely at details of the electronic structure. Figure 7 shows the energies of some molecular orbitals as functions of the applied voltage for the symmetric (a) and asymmetric (b) junctions. The straight lines are the chemical potentials, $\mu_L = E_F - eV_L$ and $\mu_R = E_F - eV_R$, of the leads. The gray area between these lines represents the energy region relevant for conduction at low (e.g. room) temperatures due to the $f_L(E) - f_R(E)$ factor in Eq. (2). We note that while for the symmetric junction the MOs behave in a symmetric way with respect to the direction of the voltage drop, the asymmetric junction shows a slight asymmetry: the HOMO on the left ($V_L < V_R$) side of Fig. 7b enters the region between the two electrochemical potentials slightly earlier (i.e. at smaller voltage) than on the right ($V_L > V_R$) side, thus leading to an earlier current increase for the negatively biased junction as compared to the positively biased one. It is encouraging to find that conduction is dominated by the HOMO, which is described better than the LUMO in the HF approximation.

---

[*] This agreement is qualitative only; in particular the gap between thresholds of current setup is much larger than observed. Indeed the HF approximation is known to overestimate the HOMO- LUMO gap (see e.g. Ref.32.). Another source of overestimating this gap in our calculation is the neglect of the electronic response of the electrodes (image interactions) to the tunneling electron.

A more important insight is obtained from comparing Figures 7 and 8. The latter shows, as functions of applied voltage, the molecular orbitals of the disconnected left and right segments of the bimolecular bridge. As expected, the symmetric junction is again characterized by a symmetric behavior. The asymmetric junction again shows a slight asymmetry in the corresponding energies, however more significant is the asymmetry in the nature of the wavefunctions. Comparing Figures 7b and 8b one sees that the HOMO of the bridge calculated as a whole (Fig. 7b) is dominated on the right side of the figure ($V_L > V_R$) by the HOMO of the right (short) molecular segment, while in the left side of the figure ($V_L < V_R$) the main contribution comes from the HOMO of the left (long) part of the molecule. This observation is consistent with the calculated overlap (not shown) between the bridge HOMO and the different atomic orbitals along the chain.

This observation implies that for a positively biased ($V_L > V_R$) junction the bridge HOMO is localized on the short (right) molecular segment while the long (left) segment constitutes an effective barrier for the motion of electrons to/from the left electrode. In the opposite case of negatively biased ($V_L < V_R$) junction the bridge HOMO is localized on the long (left) part of the molecule, and the effective barrier (for exchanging electrons with the right electrode) extends only over the short molecular segment. This is consistent with the larger current observed in the latter situation.[*] A similar mechanism for current–voltage asymmetry was first proposed in a different context in Ref. [10].

Finally, figure 9 shows electronic Mulliken populations on the groups of atoms vs. applied voltage. Only groups with marked population changes relative to equilibrium are presented. The fact that in the biased junction opposite charge is accumulated across the intermolecular gap indicates that this gap behaves as an effective capacitor. As expected, asymmetry in charge accumulation for the asymmetric junction (Fig. 7b) is observed, while no such effect is observed for the symmetric junction (Fig. 7a).

---

[*] The same transition in the character of the bridge HOMO, from dominance by the left molecular segment when the left electrode is biased negatively ($\mu_L > \mu_R$) to dominance by the right segment in the opposite case, is seen also in the symmetric junction. This, clearly, cannot lead to asymmetry in the I/V behavior of this system.

## 6. Conclusion

We have observed and discussed asymmetry in current-voltage characteristic of molecular junctions formed by bringing into contact two electrodes covered by adsorbed monomolecular layers of alkanethiol chains. Asymmetry in the current response to reversal of bias was found for asymmetric junctions characterized by different chain lengths in the two adsorbed layers. We have discussed this asymmetry in terms of simple model calculations based on the extended Hückel theory as well as calculations based on the non-equilibrium Green's function (Keldysh) method in the Hartree-Fock approximation. Both the EH and NEGF-HF calculations show asymmetry consistent with the experimental observation, however the EH calculation requires the potential distribution on the molecular bridge as an input, while the NEGF-HF calculation yields this distribution as a result of the self-consistent calculation.

Asymmetry in the molecular structure is obviously an essential attribute of rectifying molecular bridges; see Ref. [33] for a recent example. In addition, the occurrence of a potential drop along the bridge was found in the EH calculation to be an additional necessary element for observing asymmetric response. In the present system this potential distribution strongly depends on the existence of a structural weak link - the van der Waals gap between the two alkyl chains adsorbed on the two electrodes that are brought into contact. Other structures that involve a similar concept (a low conductance element connecting molecular segments of higher conductance) were recently invoked discussed .[34] Such an element essentially constitutes a molecular-scale capacitor. Obviously, the full electronic implications of such a structure in current-carrying junctions can be studied only within a non-equilibrium theory that included electron-electron interactions as was done at a simple level by the NEGH-HF calculations.

Our theoretical calculation has pointed out a reasonable mechanism for the observed rectifying action, which seems to be generic to asymmetric structures involving weak links: at opposite biases, the charge carriers are localized on molecular segments of different length, leading to different widths of the effective barrier to tunneling experienced in the two cases.

Our theoretical results are in qualitative agreement with the experimental observations with respect to both the magnitude and the sign of the effect, however the is a large quantitative discrepancy, mostly reflected in the conduction gap that is strongly overestimated by the theoretical model. This disagreement arises from several sources. First the theoretical calculation focuses on a single molecular bridge while the experimental system deals with a molecular layer of relative large lateral extent. Secondly, HF theory is known to strongly overestimate the HOMO-LUMO gap, in particular overestimating the energies of LUMO levels. Finally, the electronic response of the metal to the tunneling electron (image interactions) was not taken into account in the present calculation. Such interaction will lower the energy of a transient molecular ion, thus decreasing the conduction gap.

In spite of its inherent simplicity, our model calculation does appear to account for the observed asymmetric behavior of asymmetric molecular junctions characterized by weak structural links. The existence of high resistance elements within otherwise low resistance molecular segments makes it possible to change independently the electrostatic state of different parts of a molecular system. Such internal molecular capacitors therefore constitute a potentially convenient way to control the performance of molecular junctions, and further studies of this concept and its implications seem to be worthwhile.

**Appendix A: Large systems at HF level**

The main complexity in the NEGF calculations for large molecular systems stems from the need to evaluate and store Green's function (GF) matrices for many energies, required for the evaluation of integrals such as (9). Depending on the structure of the integrand (e.g. number and sharpness of resonance, we may encounter the need to do such calculations on a fine grid on the energy axis. At each energy point we need to evaluate the GF (inversion of a big matrix) and sometimes store it, implying a big demand on CPU time and memory. Here we present an approach that overcomes both limitations and can be used for NEGF calculations of big systems. The method described below is valid in cases where the self-energies arising from the inter-particle interactions are only of the irregular type[*] (as is the case at the HF level of approximation) and the retarded self-energies due to the electrodes can be assumed to be energy independent; the so called wide-band limit of the external reservoir (see, e.g. Ref. [11]).

---

[*] This implies the vanishing of the lesser and greater self energies and (for steady state situations) the energy-independence of the retarded (advanced) self-energies; see Refs 35. and 36.

*Hamiltonian and Green functions.* The reduced Hamiltonian for the molecular bridge, obtained by projecting out the electrode sub-spaces, may be written as in the basis of $N$ Löwdin orbitals in the form

$$\mathbf{H} = \mathbf{H}_0 + \mathbf{\Sigma}^r \tag{15}$$

where $\mathbf{H}_0$ is the zero order Hamiltonian (the first two terms of Eq. (1)) and $\mathbf{\Sigma}^r$ is the retarded self-energy that contains contributions from the molecule-electrode coupling as well as the from the electron-electron interaction (calculated in the HF approximation). The Hamiltonian (15) can be diagonalized using

$$\mathbf{HV} = \mathbf{VH}_d \quad \Leftrightarrow \quad \mathbf{H} = \mathbf{VH}_d \mathbf{V}^{-1} \tag{16}$$

where $\mathbf{V}$ is matrix formed from columns of the right eigenvectors of the matrix $\mathbf{H}$ and $\mathbf{H}_d$ is a diagonal matrix of complex eigenvalues $\{E_n\}$, (n = 1, ..., N) of $\mathbf{H}$. The retarded Green's function satisfies (cf. Eq. (10))

$$\mathbf{G}^r(E) = [E - \mathbf{H}]^{-1} = \mathbf{V}[E - \mathbf{H}_d]^{-1} \mathbf{V}^{-1} \tag{17}$$

or

$$G_{ij}^r(E) = \sum_l V_{il} \frac{1}{E - E_l} (V^{-1})_{lj} \tag{18}$$

The poles $E_l$ satisfy $E_l = \varepsilon_l - i\gamma_l$ and $\gamma_l > 0$.

The matrix elements of the advanced Green's function, $G^a(E) = \left[G^r(E)\right]^\dagger$, are given by

$$G_{ij}^a(E) = \sum_l (V^{-1})_{li}^* \frac{1}{E - E_l^*} V_{jl}^* \tag{19}$$

The corresponding lesser and greater Green's functions can be obtained from the Keldysh equation, Eq. (8). The result is

$$G_{ij}^{<}(E) = i\sum_{p,r} V_{ip}\left(M_{pr}^{L}f_{L}(E) + M_{pr}^{R}f_{R}(E)\right)V_{jr}^{*}$$
$$\times \frac{1}{E - \varepsilon_p + i\gamma_p} \cdot \frac{1}{E - \varepsilon_r - i\gamma_r} \quad (20)$$

$$G_{ij}^{>}(E) = -i\sum_{p,r} V_{ip}\left(M_{pr}^{L}[1 - f_{L}(E)]\right.$$
$$\left. + M_{pr}^{R}[1 - f_{R}(E)]\right)V_{jr}^{*}$$
$$\times \frac{1}{E - \varepsilon_p + i\gamma_p} \cdot \frac{1}{E - \varepsilon_r - i\gamma_r} \quad (21)$$

where

$$M_{pr}^{K} = \sum_{s,t} \left(V^{-1}\right)_{ps} \Gamma_{st}^{K} \left(V^{-1}\right)_{rt}^{*} \quad ; \quad K = L, R \quad (22)$$

and the matrix $\Gamma$ is defined by Eq. (3).

*The density matrix.* The main numerical problem we face is the need to do the energy integrals that appear in many of expressions of interest, e.g. Eq. (9) for density matrix. In the general case such integral have to be done numerically and this constitutes the main numerical obstacle in the application of the NEGF to practical problems. Under our assumptions the retarded self-energy was assumed energy independent, while the energy dependence of the lesser and greater self-energies is known (cf. Eqs. (20), (21)). Using Eq. (20) in (9) leads to

$$\rho_{ij} = \sum_{K=L,R} \sum_{p,r} \left\{ V_{ip} M_{pr}^{K} I_{pr}^{K} V_{jr}^{*} \right\} \quad (23)$$

$$I_{pr}^{K} = \int_{-\infty}^{+\infty} \frac{dE}{2\pi} \frac{1}{E - \varepsilon_p + i\gamma_p} \frac{1}{E - \varepsilon_r - i\gamma_r} f_{K}(E) \quad (24)$$

This integral can be evaluated exactly as described below.

*Integral evaluation.* Consider the integral

$$I = \int_{-\infty}^{+\infty} \frac{dE}{2\pi} \frac{1}{E-\varepsilon_1+i\gamma_1} \cdot \frac{1}{E-\varepsilon_2-i\gamma_2} \cdot \frac{1}{\exp[\beta(E-\mu)]+1}$$

$$= \frac{\beta}{2\pi} \int_{-\infty}^{+\infty} dx \frac{1}{x-\beta(\varepsilon_1-\mu-i\gamma_1)} \cdot \frac{1}{x-\beta(\varepsilon_2-\mu+i\gamma_2)} \cdot \frac{1}{e^x+1}$$

(25)

(the last equality is obtained by using $x = \beta(E-\mu)$). This integral can be evaluated by contour integration. The integrand has simple poles at (see Fig. 10)

$$\begin{aligned} x = x_1 &\equiv \beta(\varepsilon_1-\mu-i\gamma_1) \\ x = x_2 &\equiv \beta(\varepsilon_2-\mu+i\gamma_2) \\ x = y_n &\equiv i(\pi+2\pi n), \quad n=0,\pm 1,\ldots \end{aligned}$$

(26)

One can close the contour of integration either along the $C_1$ or the $C_2$ contours (Fig. 10). We get accordingly

$$I = \begin{cases} i\beta \left[ \dfrac{1}{x_2-x_1} \dfrac{1}{e^{x_2}+1} - \sum_{n=0}^{\infty} \dfrac{1}{(y_n-x_1)(y_n-x_2)} \right]; & C_1 \\ i\beta \left[ \dfrac{1}{x_2-x_1} \dfrac{1}{e^{x_1}+1} - \sum_{n=0}^{\infty} \dfrac{1}{(y_n+x_1)(y_n+x_2)} \right]; & C_2 \end{cases}$$

(27)

Using [37]

$$\sum_{n=0}^{\infty} \frac{1}{(n+a)(n+b)} = \frac{1}{b-a}\{\psi(b)-\psi(a)\}$$

(28)

where

$$\begin{aligned} \psi(z) &= \frac{\Gamma'(z)}{\Gamma(z)} \quad \text{Psi (digamma) function} \\ \Gamma(z) &= \int_0^{\infty} dt\, t^{z-1} e^{-t} \quad \text{Gamma function} \end{aligned}$$

(29)

we obtain

$$I = \begin{cases} \dfrac{i\beta}{x_2-x_1} \left[ \dfrac{1}{e^{x_2}+1} - \dfrac{i}{2\pi}\left\{\psi\left(\dfrac{1}{2}+i\dfrac{x_2}{2\pi}\right)-\psi\left(\dfrac{1}{2}+i\dfrac{x_1}{2\pi}\right)\right\} \right]; & C_1 \\ \dfrac{i\beta}{x_2-x_1} \left[ \dfrac{1}{e^{x_1}+1} - \dfrac{i}{2\pi}\left\{\psi\left(\dfrac{1}{2}-i\dfrac{x_2}{2\pi}\right)-\psi\left(\dfrac{1}{2}-i\dfrac{x_1}{2\pi}\right)\right\} \right]; & C_2 \end{cases}$$

(30)

One can show that the two results are equal using[38]

$$\psi(1-z) = \psi(z) + \pi \cot(\pi z) \tag{31}$$

Numerical evaluation of the Psi (digamma) function is, however troublesome. So instead in our numerical calculation we use the following simple scheme for evaluation of the infinite sum entering the integral expression (27). For either sum in (27) let $S_N = \sum_{n=0}^{N} \ldots$ and choose $N$ large enough so that the rest of the sum may be presented as an integral. This leads to

$$\sum_{n=0}^{\infty} \frac{1}{(y_n \mp x_1)(y_n \mp x_2)} = S_N - \frac{1}{(2\pi)^2} \left\{ \frac{1}{N+1} - \frac{1 \pm i\frac{x_1+x_2}{2\pi}}{2(N+1)^2} \right\} \tag{32}$$

The evaluation of I is then reduced to summation to obtain $S_N$, an easy numerical task.

**Acknowledgements**.  MM acknowledges the donors of the Petroleum Research Fund, administered by the ACS for partial support of this research.  Additional support was provided at UCB by the National Science Foundation (CHE-0079225). Research at TAU was supported by the U.S. - Israel binational Science Foundation, by the Israel National Science Foundation and by the Israel Ministry of Science.

**Figure captions**

**Figure 1.** (a) Schematic view of a Hg-Au alkanethiol bilayer junction. (b) A single molecule model used in the calculations.

**Figure 2.** A representative set of three types of voltammetric curves observed with decanethiol monolayer coated gold substrates in a 1 mM $Ru(NH_3)_6^{3+}$, 0.5 M KCl solution.

**Figure 3**. Two $I/V$ curves recorded at 500 V/s for the Hg-Au junctions described in the inset. The $I/V$ curves are printed following a constant capacitive background subtraction. Negative bias refers to the negative voltage applied to the Hg side of the junction.

**Figure 4**. Potential profiles used in the EH calculations. See text for details.

**Figure 5**. Current vs. applied voltage obtained from an extended Hückel calculation for a symmetric ($n=m=3$; dashed line) and asymmetric ($n=5$, $m=1$; solid line) junctions for models A, B, and C of the potential distribution across the junction (see Fig. 4).

**Figure 6**. (a) Current vs. applied voltage obtained from a NEGF-HF calculation for the symmetric ($n=m=3$; dashed line) and asymmetric ($n=5$, $m=1$; solid line) junctions. (Here and in the following figure the circles and triangles denote the voltage points were actual calculations were made). (b) Similar results computed for 8-carbon junctions of variable asymmetry.

**Figure 7**. Molecular orbitals of the entire bimolecular bridge vs. applied voltage for the (a) symmetric and (b) asymmetric junctions. Shown are the HOMO (solid line) and two MOs immediately below it (dashed and dotted lines) as well as the LUMO (solid line)

and two MOs immediately above it (dashed and dotted lines). The circles denote the voltage points for which calculations were made; the lines interpolate between these points. Also shown are the electrochemical potentials (straight lines) of the electrodes.

**Figure 8**. Molecular orbitals of the individual molecular segments of the bridge vs. applied voltage for the (a) symmetric ($n=m=3$) and (b) asymmetric ($n=5$, $m=1$) junction. Shown are HOMO with the level immediately below it and the LUMO with the level immediately above it for the left (solid lines, circles) and right (dashed lines, triangles) molecular segments. Also shown are the chemical potentials of the electrodes (straight lines). The circles and triangles denote voltage points were actual calculations were done.

**Figure 9**. Mulliken electron populations on different atomic groups (C and connected H atoms) vs. applied voltage for the (a) symmetric ($n=m=3$) and (b) asymmetric ($n=5$, $m=1$) junction. Only groups with marked population changes relative to equilibrium are presented.

**Figure 10**. Contours and poles in the complex plane. See text for details.

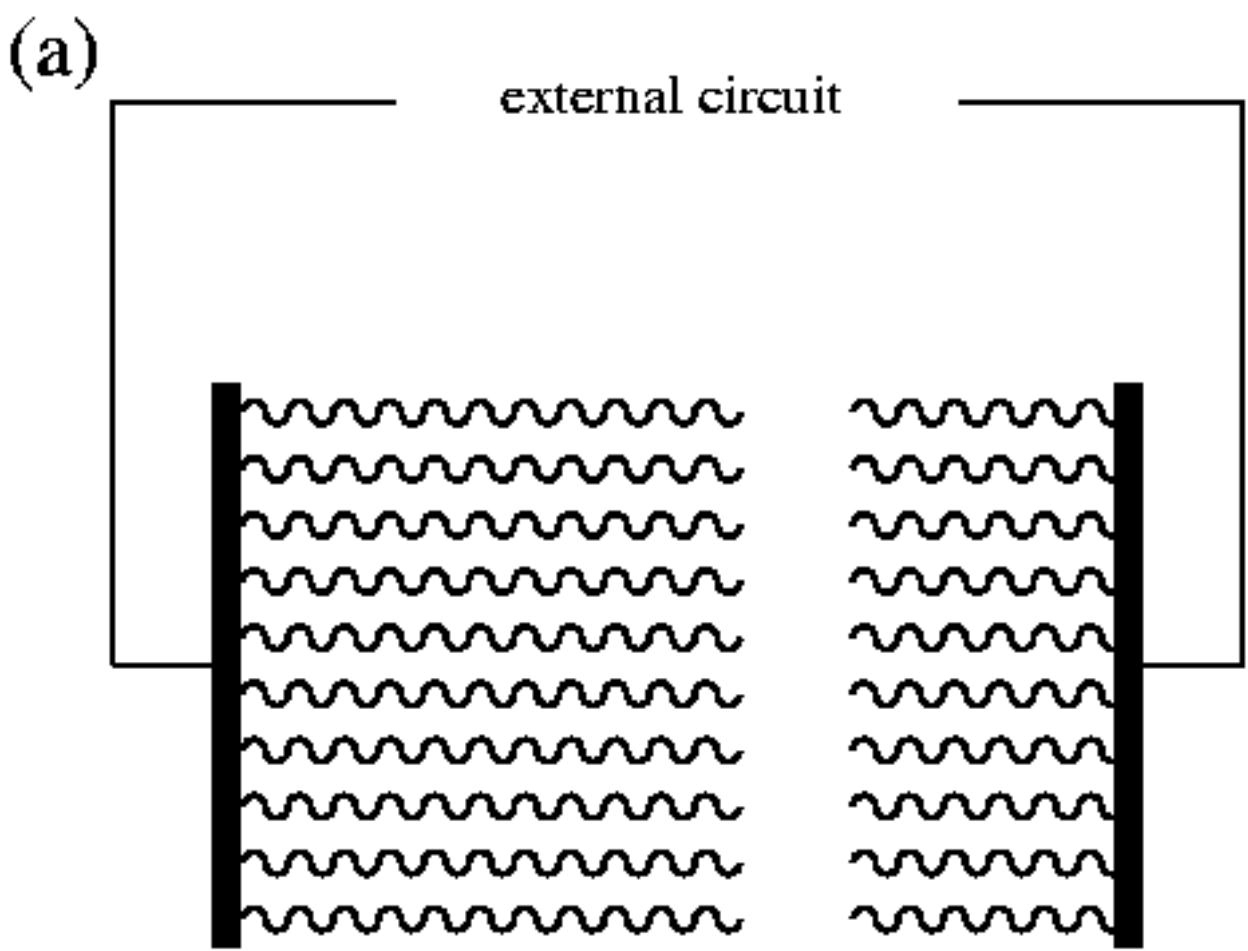
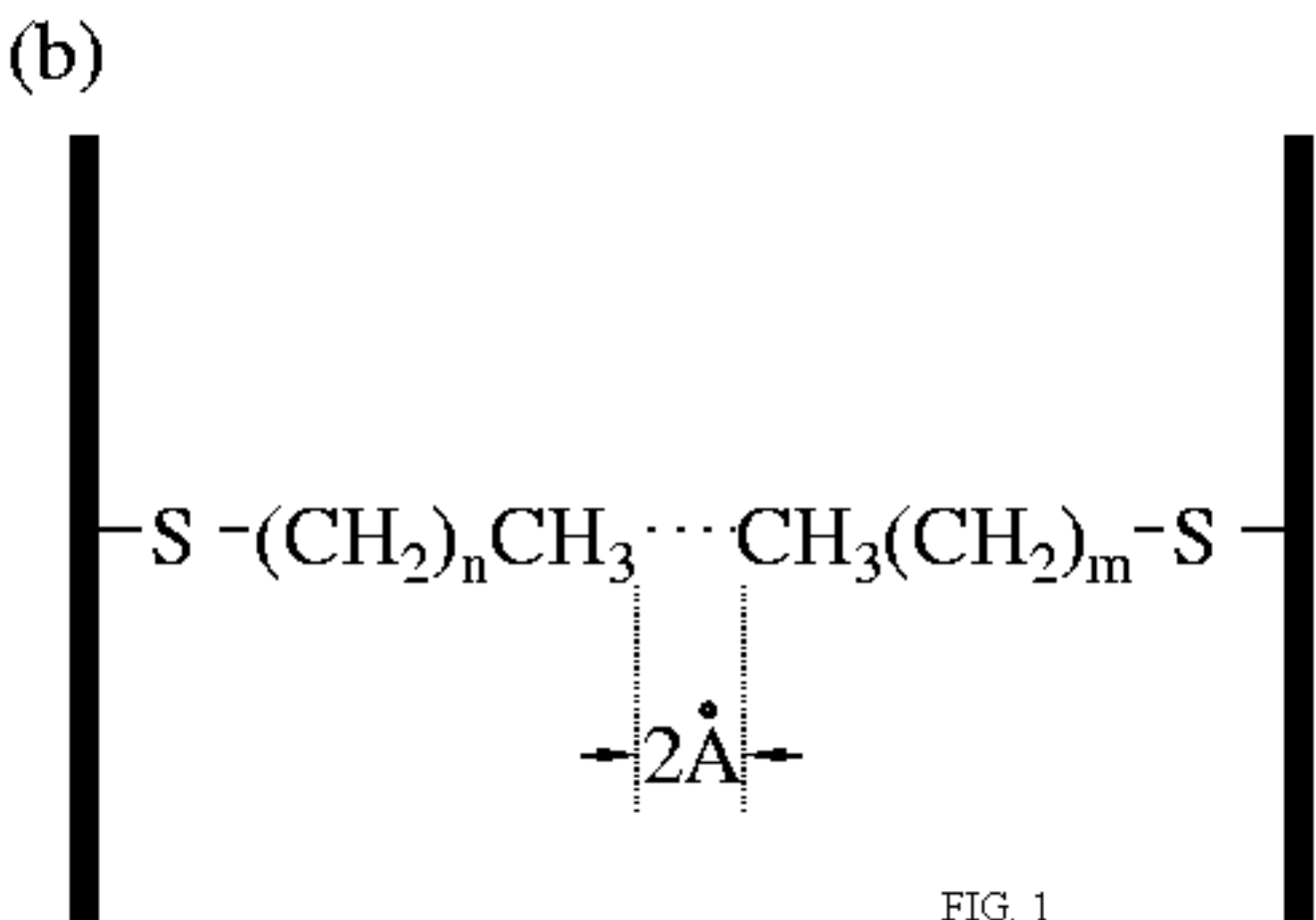

FIG. 1

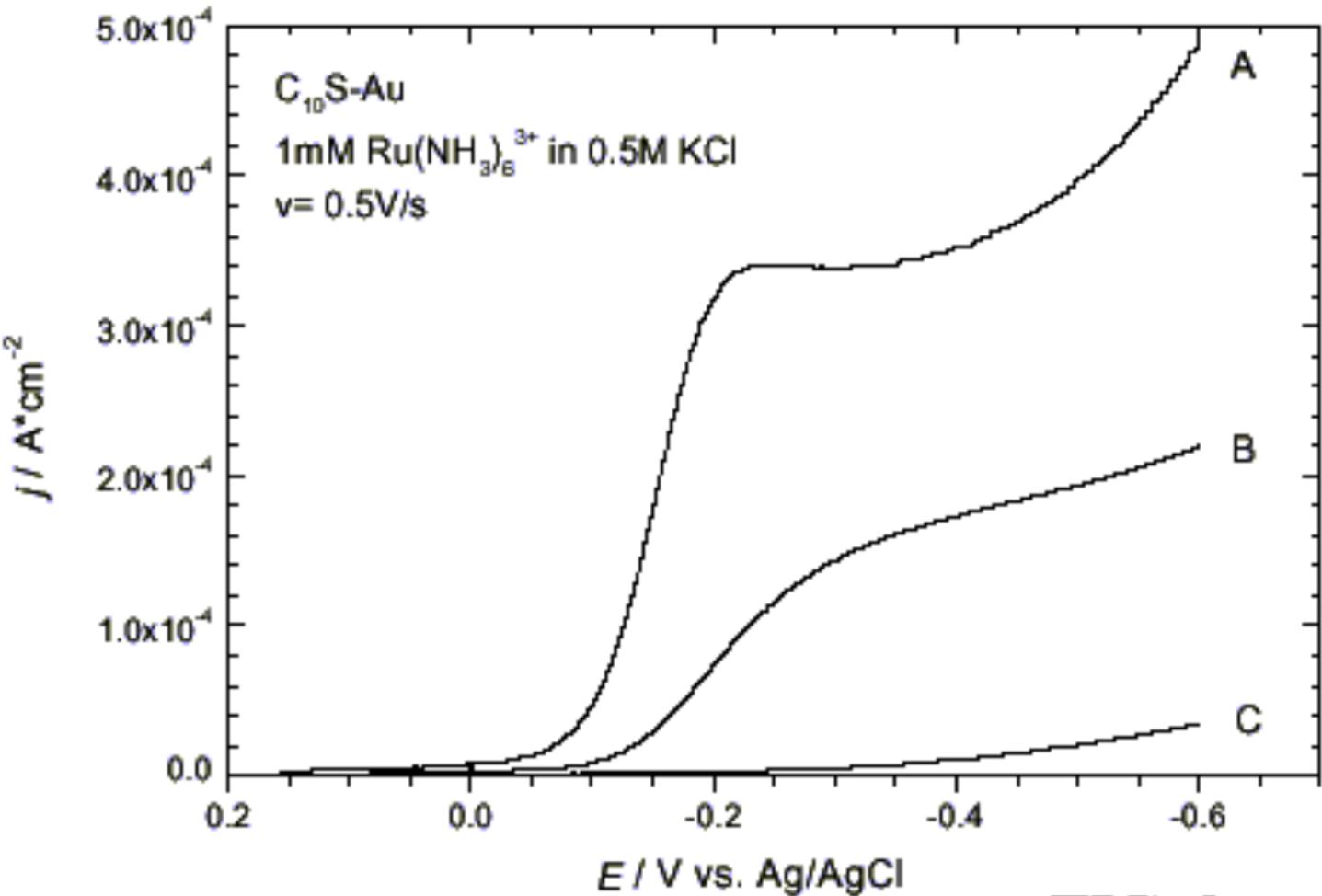

**FIG. 2**

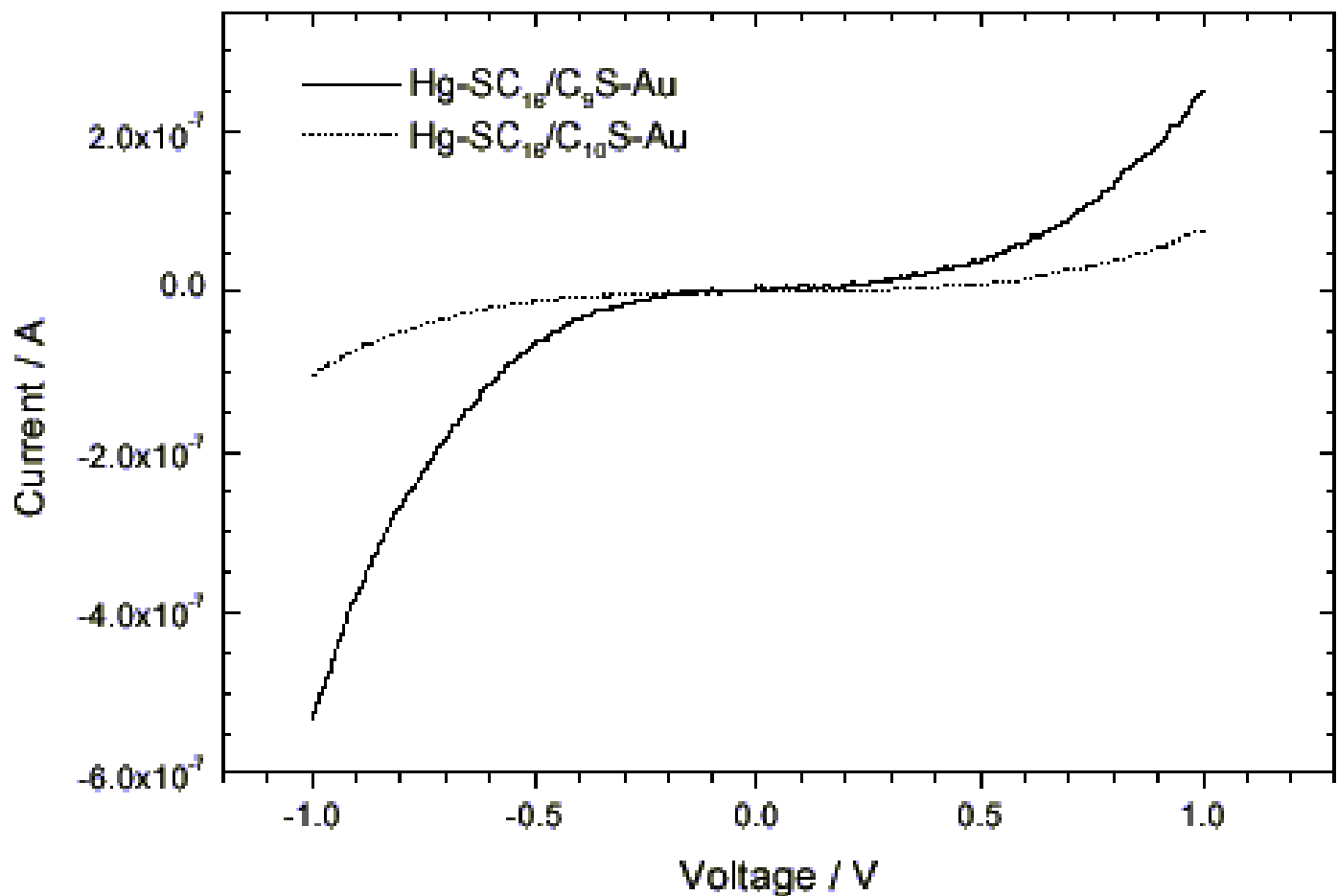

FIG 3

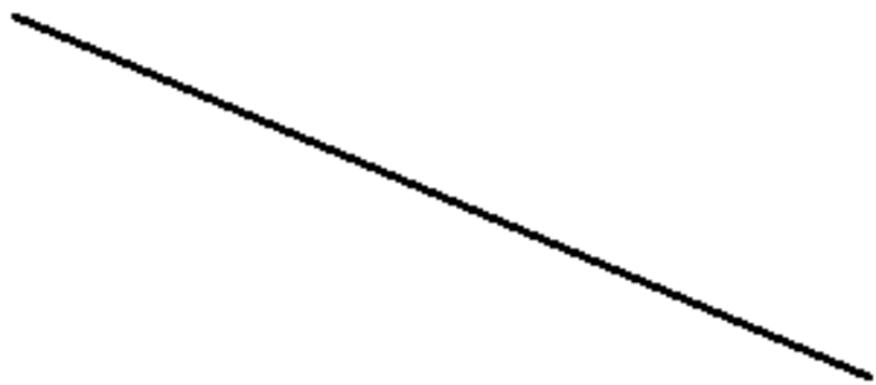
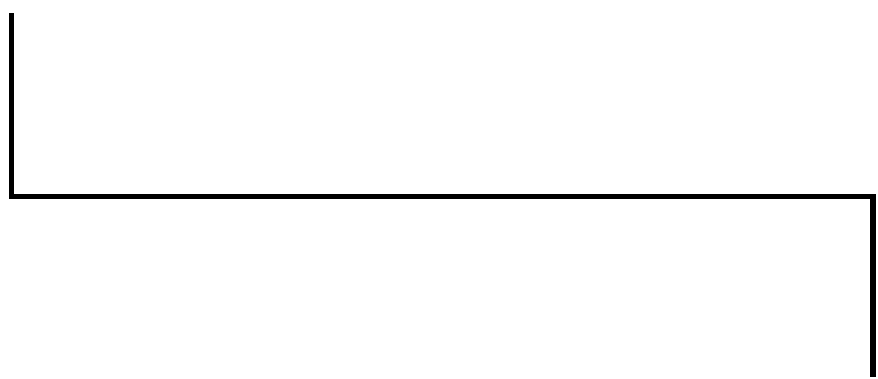
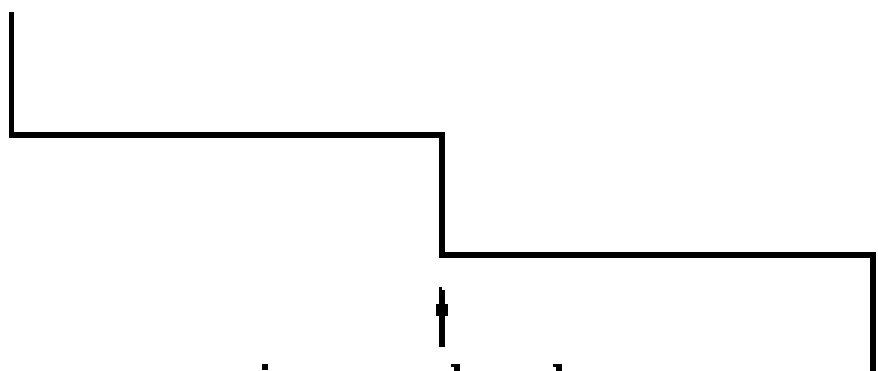

FIG. 4    intermolecular gap

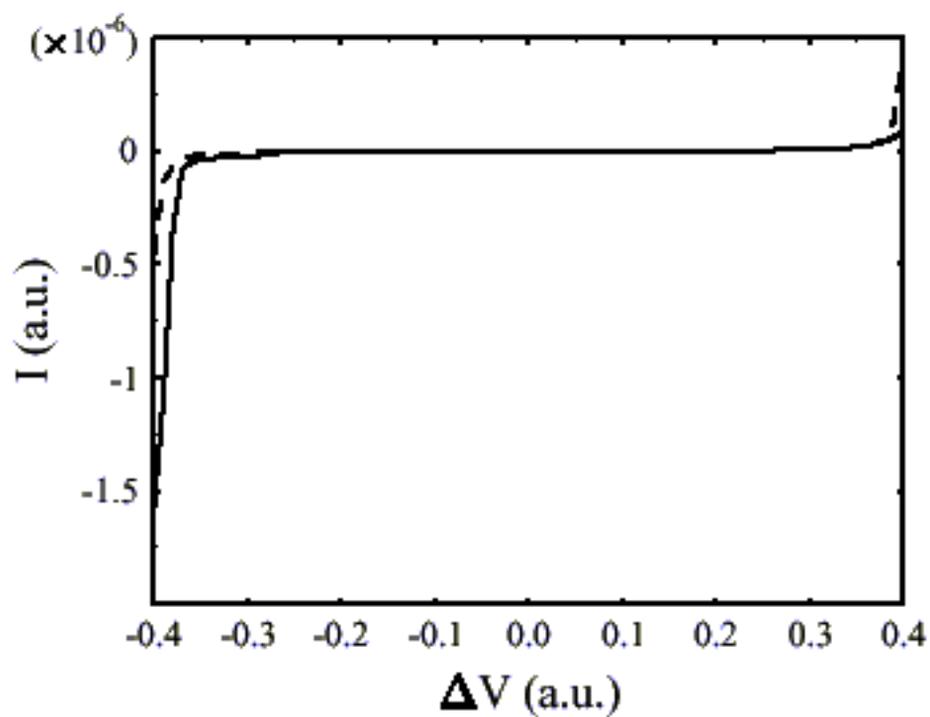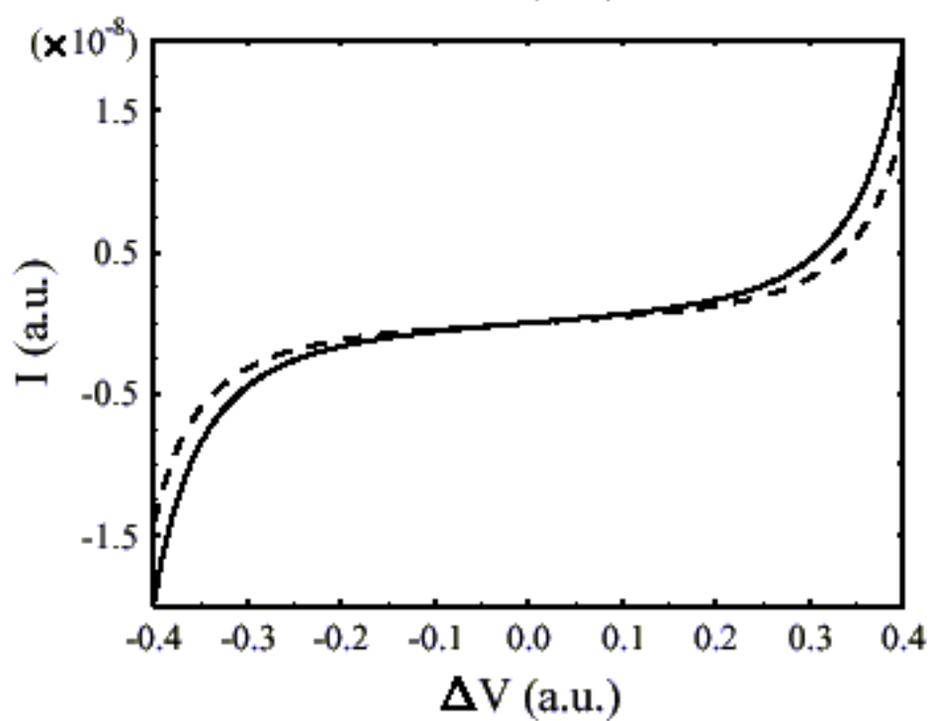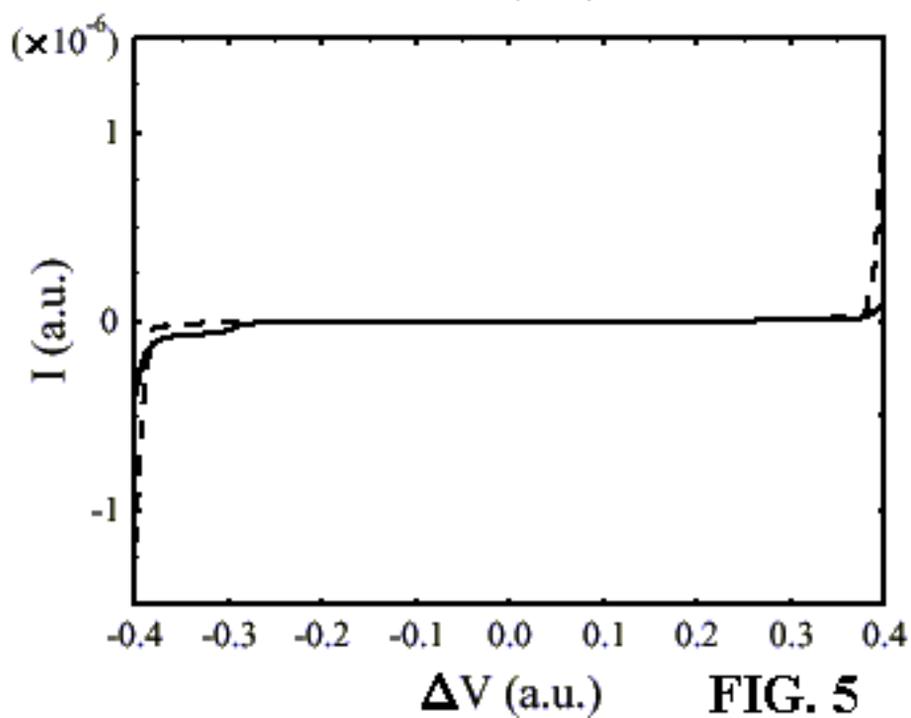

FIG. 5

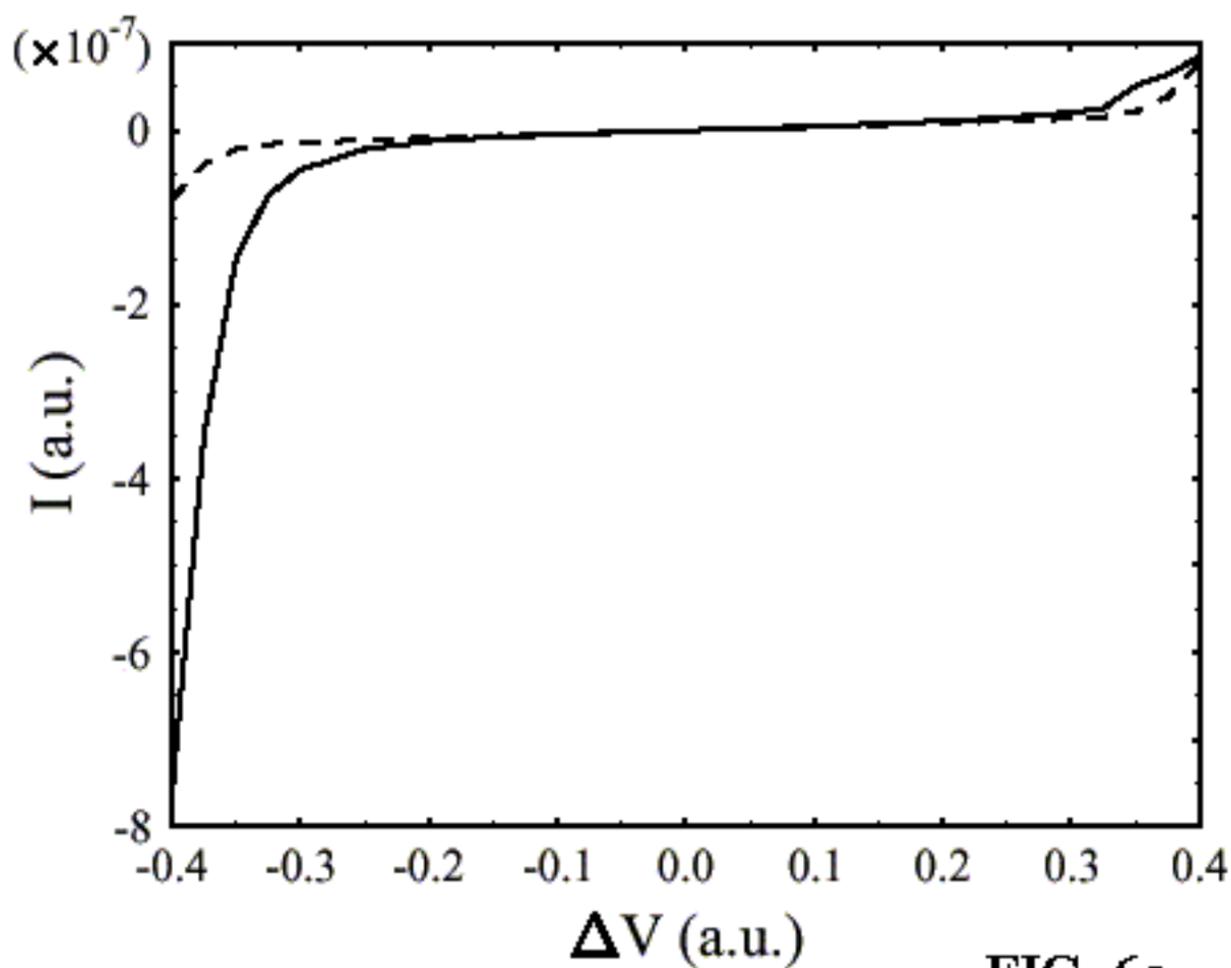

FIG. 6a

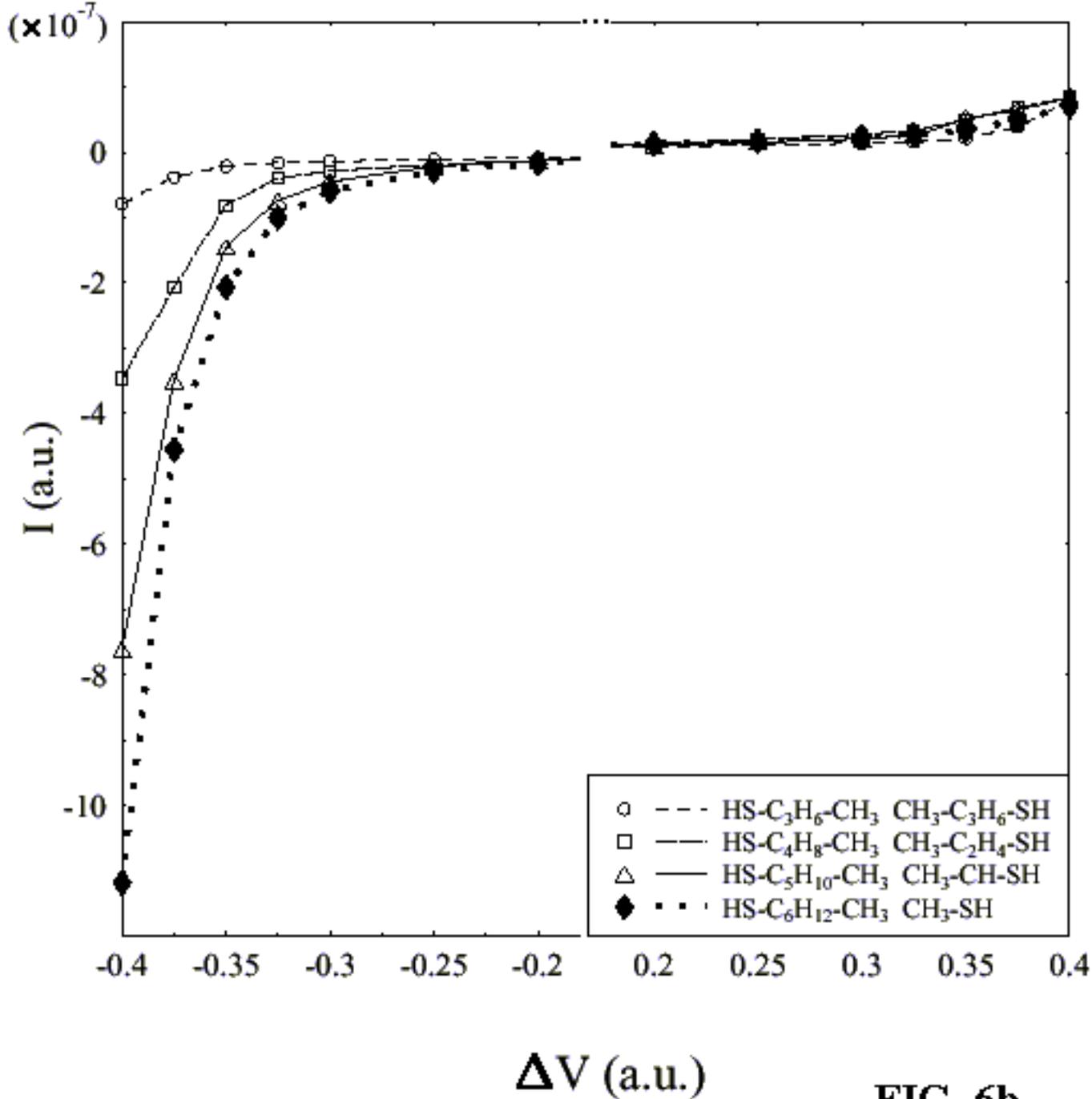

FIG. 6b

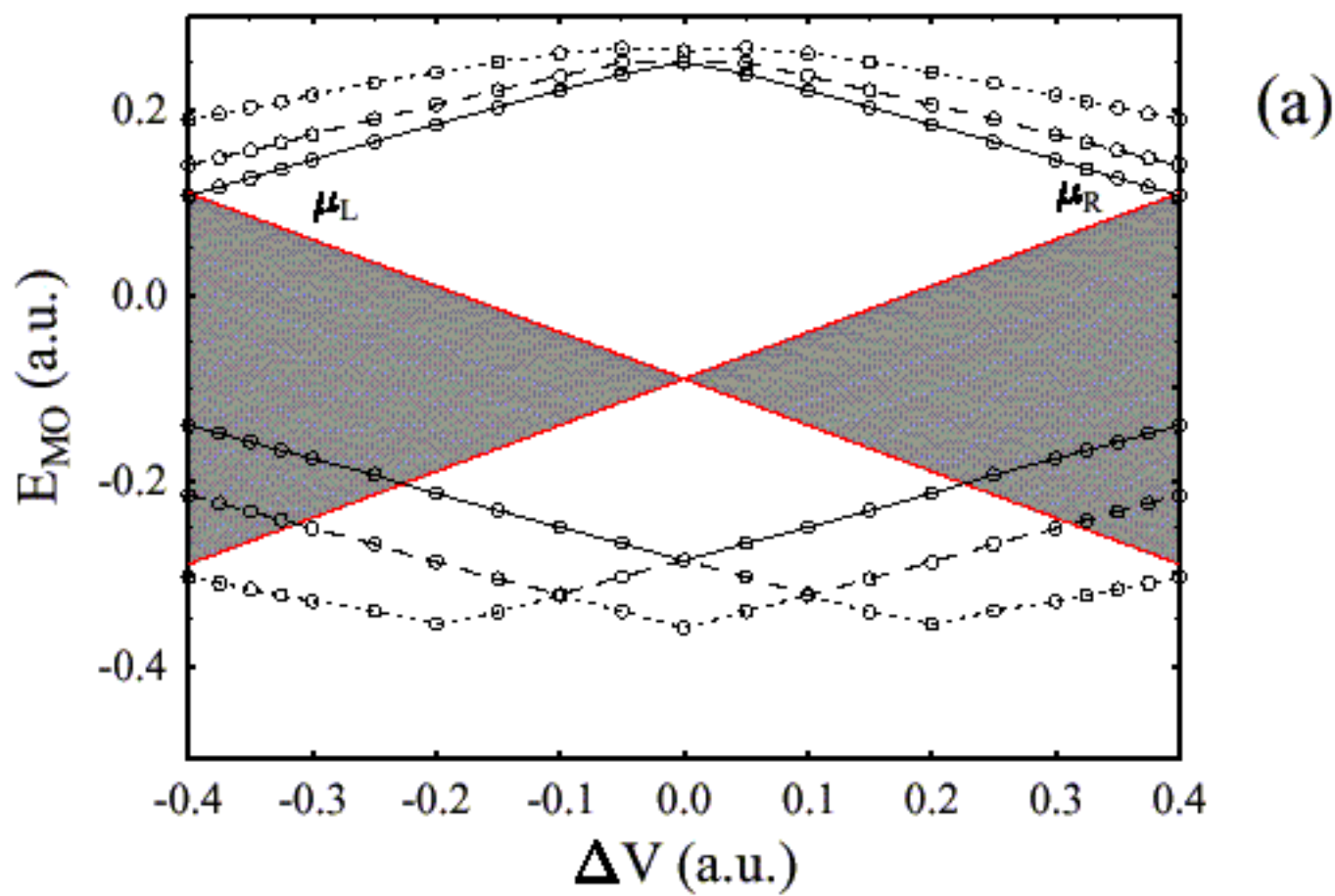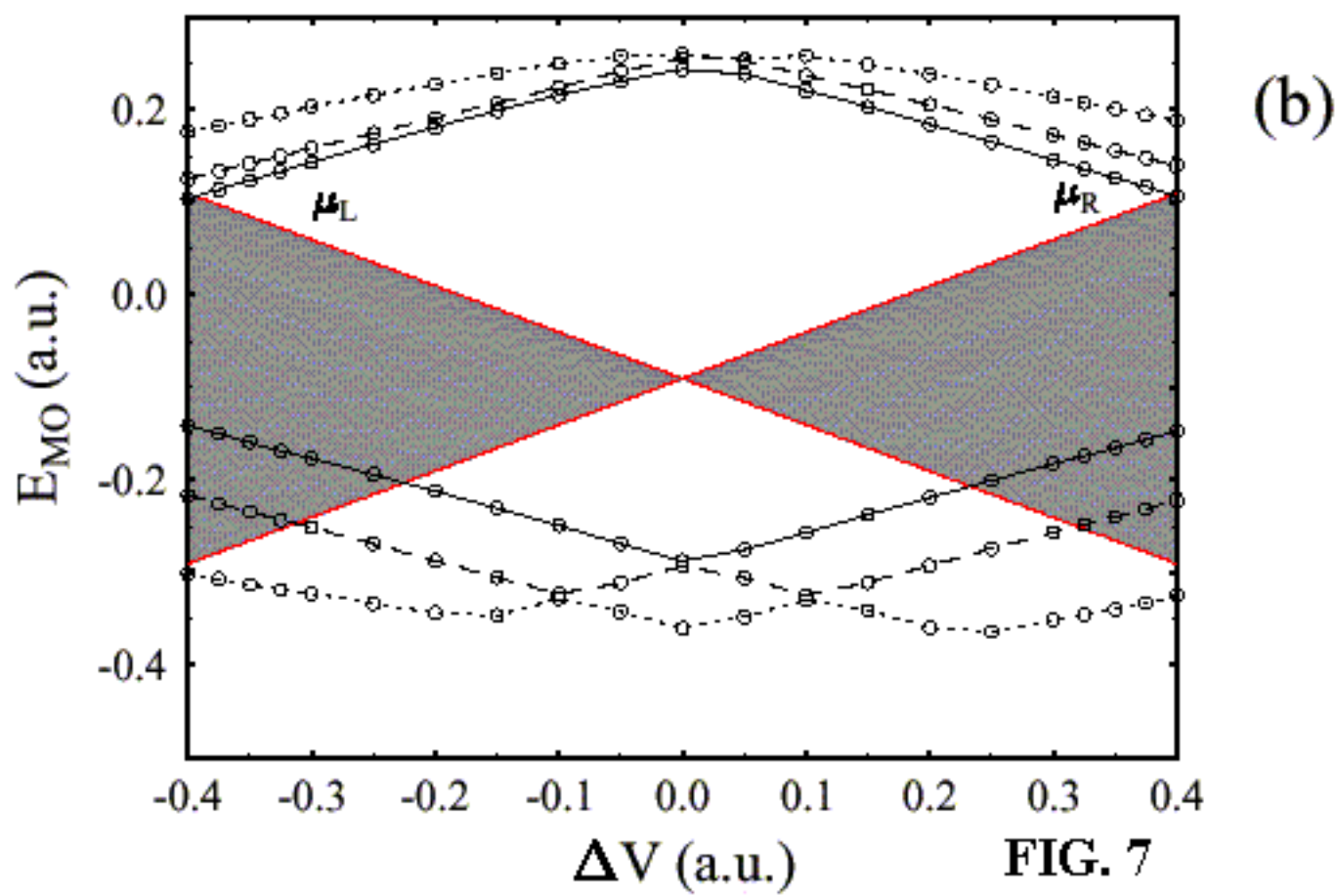

FIG. 7

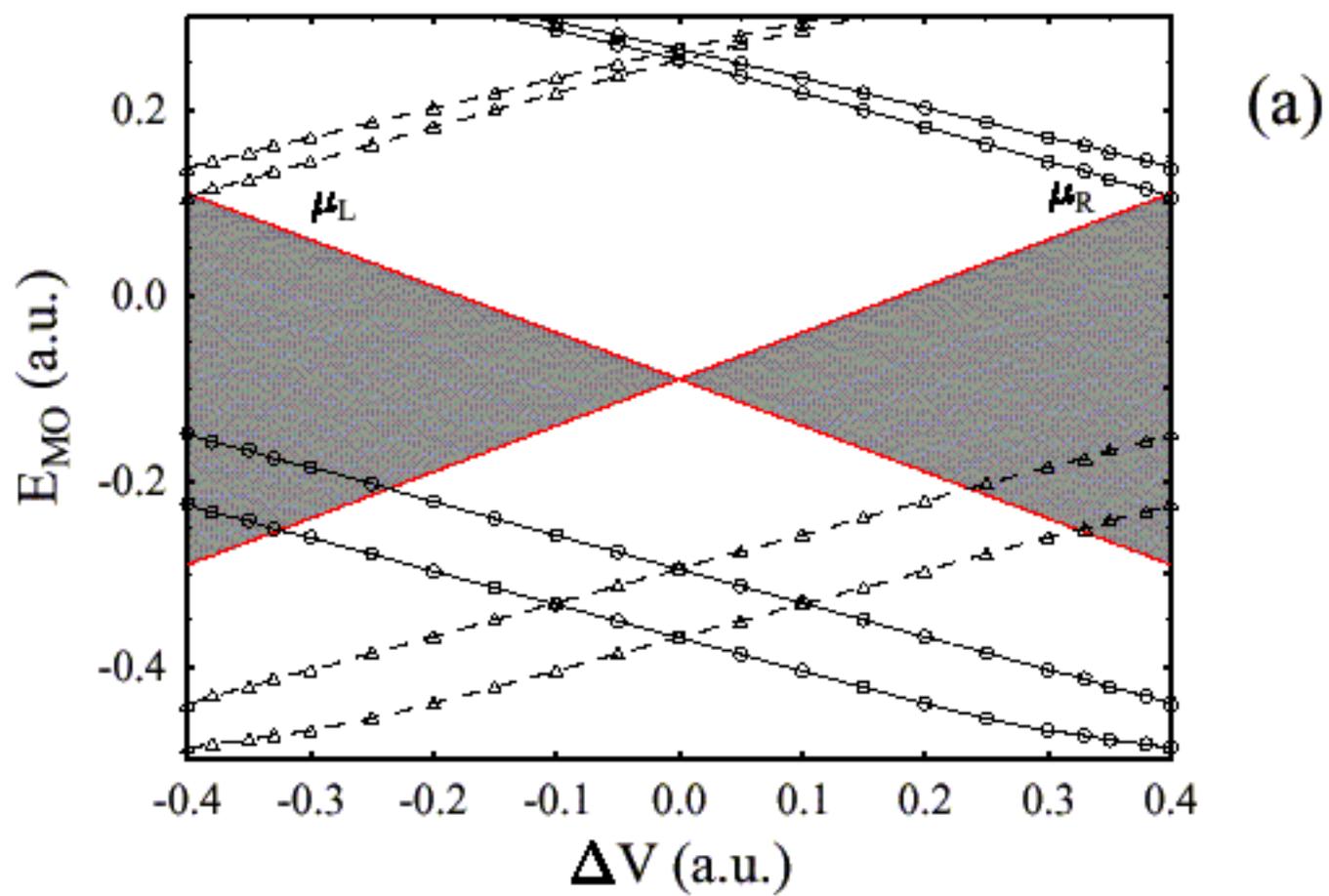

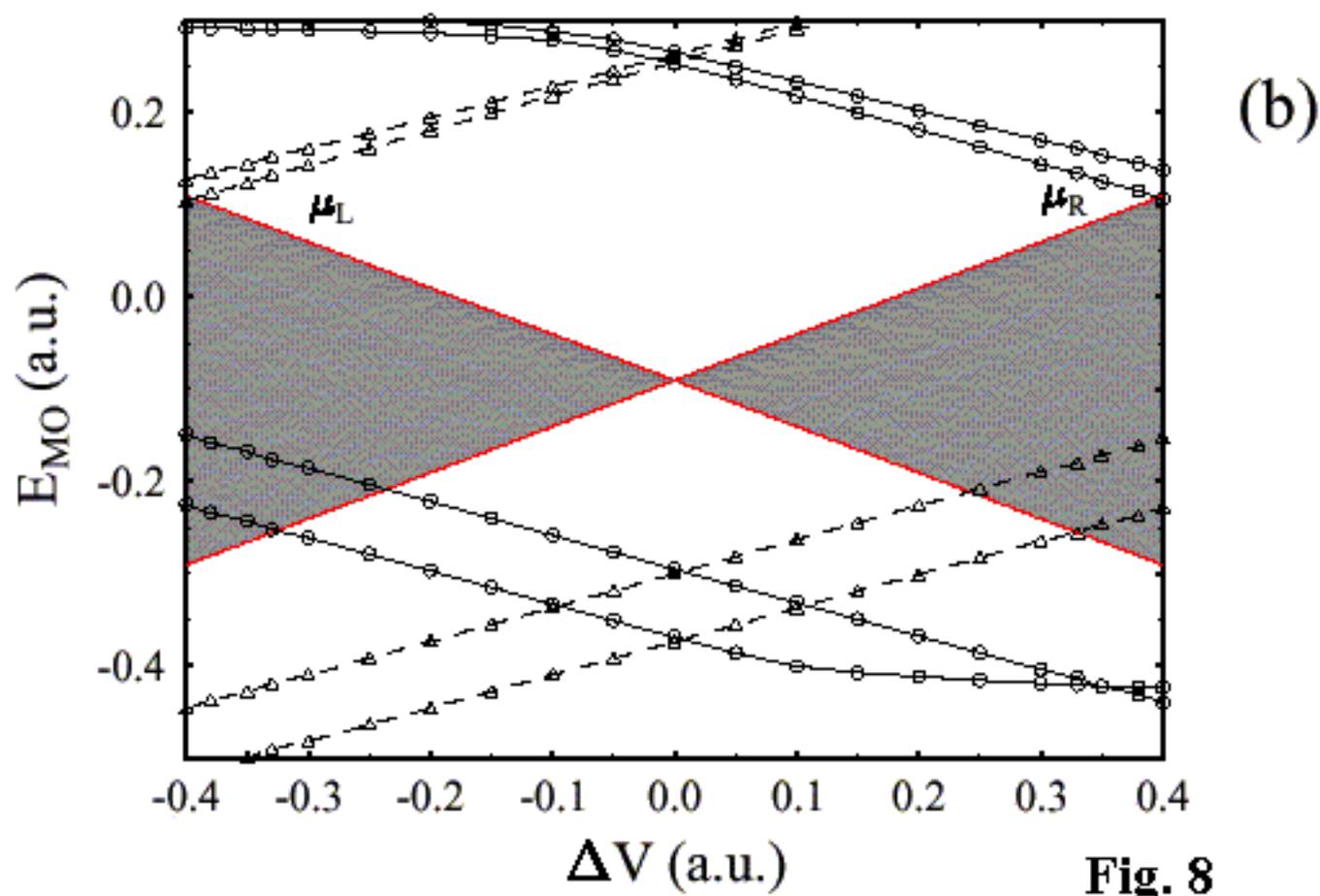

Fig. 8

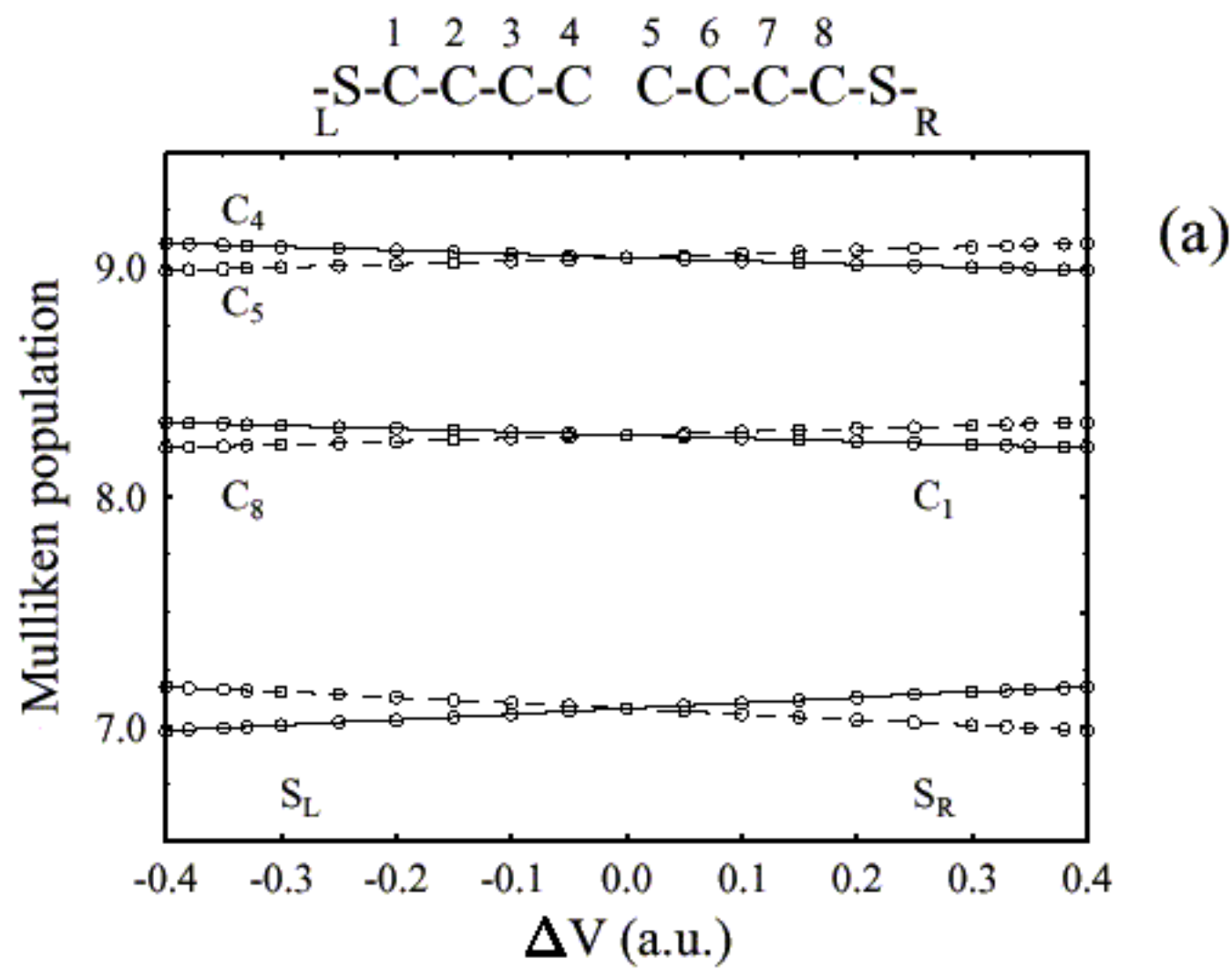
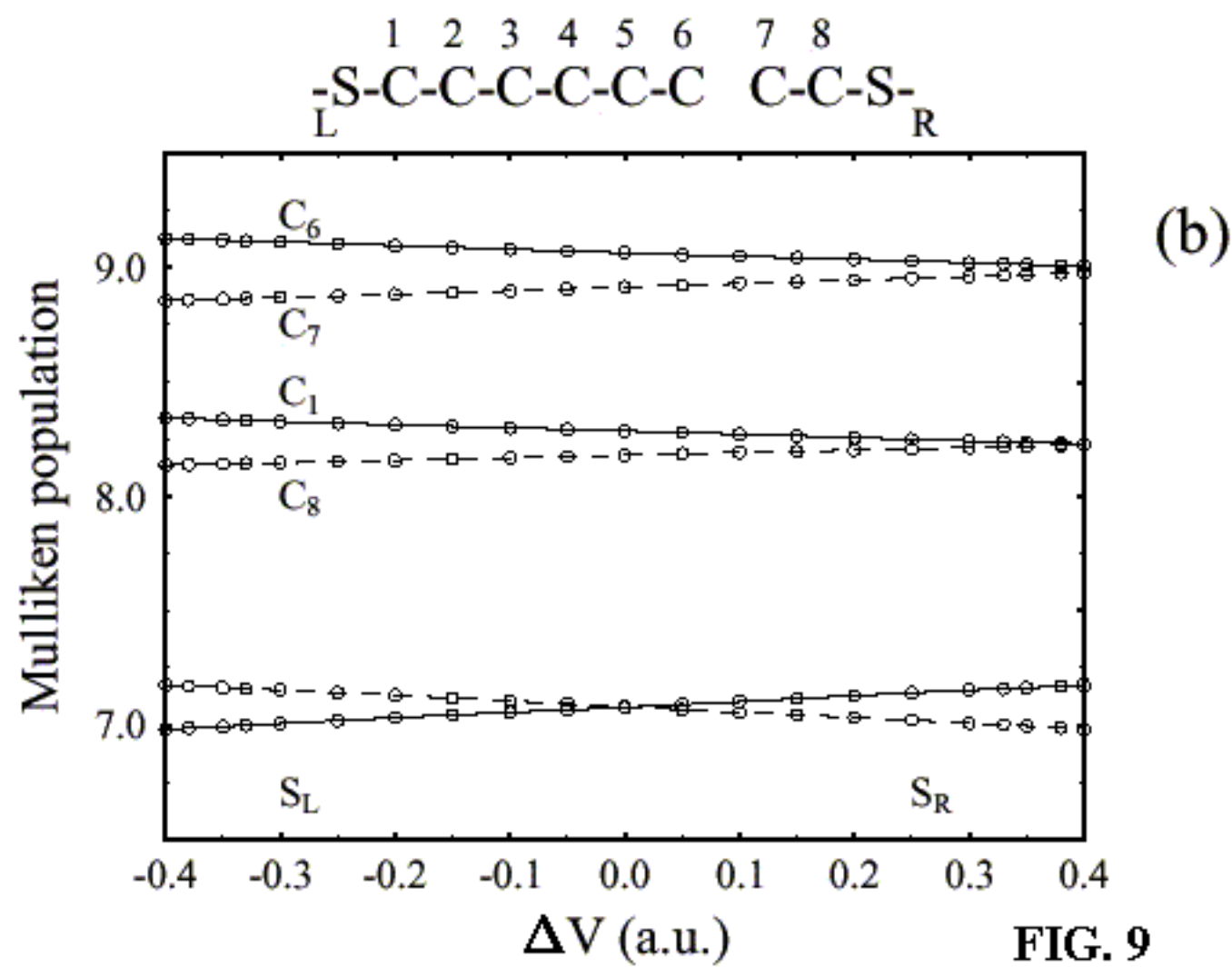

FIG. 9

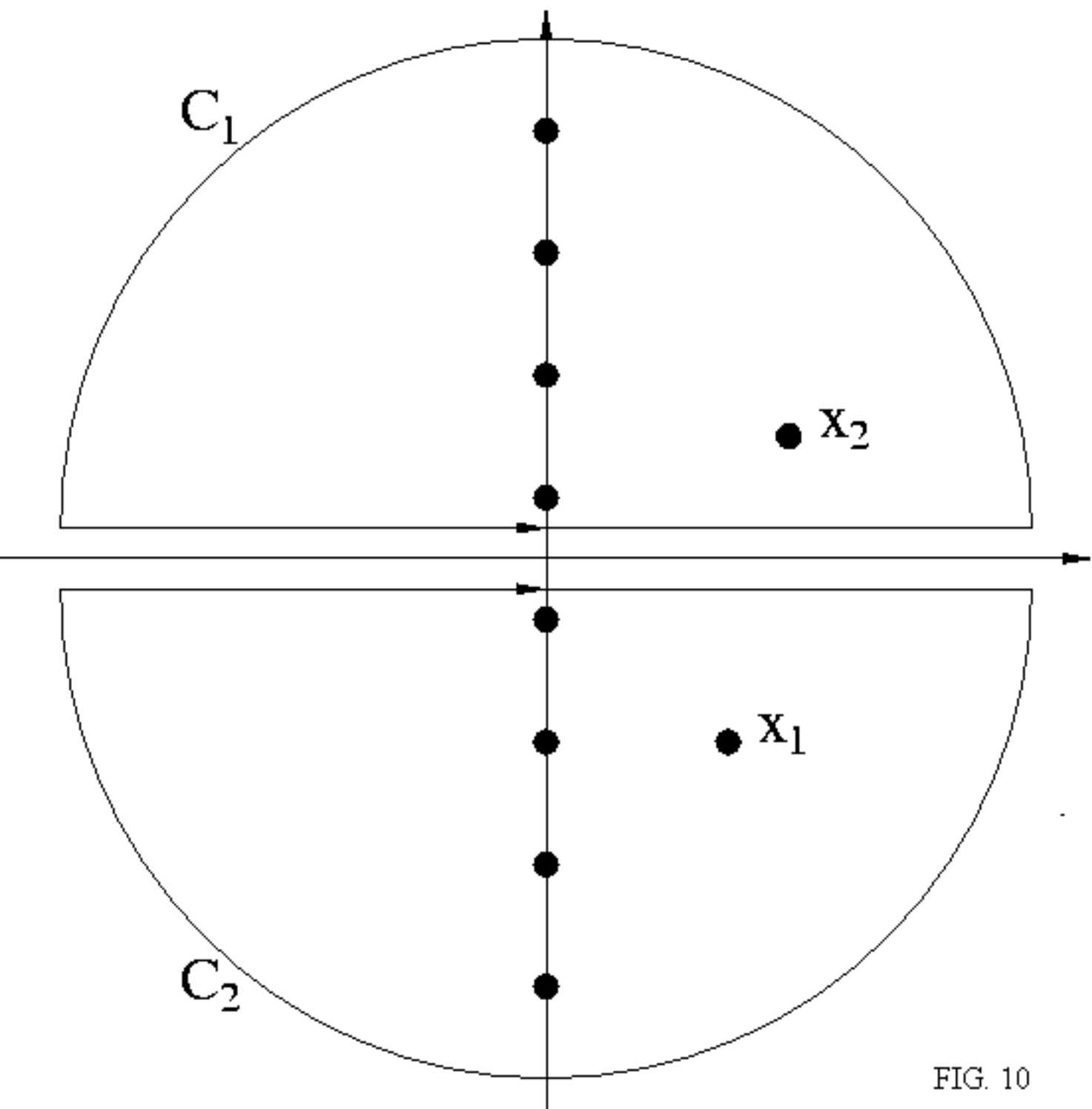

FIG. 10